\documentclass[aps,prl,twocolumn,groupedaddress]{revtex4}
\usepackage{graphicx}
\begin{document}

\title{Slow dynamics and stress relaxation in a liquid as an elastic medium}
\author{Kostya Trachenko}
\address{Department of Earth Sciences, University of Cambridge,
Downing Street, Cambridge, CB2~3EQ, UK}

\begin{abstract}
We propose a new framework to discuss the transition from exponential relaxation in a liquid to the regime of slow dynamics.
For the purposes of stress relaxation, we show that a liquid can be treated as an elastic medium. We discuss that, on
lowering the temperature, the feed-forward interaction mechanism between local relaxation events becomes operative, and
results in slow relaxation.
\end{abstract}


\maketitle

At high temperature, a liquid under external perturbation relaxes exponentially fast: a relaxing quantity $q(t)\propto
\exp(-t/\tau)$, where $\tau$ is associated with the transition over a single activation barrier. This is known as Debye
relaxation. On lowering the temperature, relaxation becomes qualitatively different. In many cases, it is well approximated
by the stretched-exponential function, $q(t)\propto \exp(-(t/\tau)^\beta)$, where $0<\beta<1$; smaller $\beta$ corresponds
to slower relaxation. Slow relaxation is universally seen in many supercooled liquids, glasses and other disordered systems
\cite{dyre,ngai,phillips}. It describes a very sluggish dynamics: in the wide data range, it decays as a logarithm of time.
The transition to slow relaxation signals the onset of glass transformation range \cite{dyre}, which ends with the dynamical
arrest at $T_g$, glass transition temperature. $T_g$ is conventionally defined from the condition that relaxation time
$\tau$ exceeds the experimental time scale of 100--1000 seconds. Slow relaxation can be seen below $T_g$ as well, although
stronger perturbation (e.g. large pressure \cite{bra}) is needed to induce relaxation.

The universal character of slow relaxation strongly suggests that there should be some fundamental slowing down mechanism,
that kicks in on lowering the temperature. Since stretched-exponential relaxation (SER) was introduced by Kohlrausch in 1854
\cite{kohl}, it has proved difficult to rationalize it from the first principles, without invoking postulates or
assumptions. This has contributed to the lack of consensus about its physical origin \cite{zwanzig,phil1}. For a more
detailed discussion, the reader is referred to papers that review slow relaxation \cite{dyre,ngai,phillips}.

Our thinking of the liquids is shaped by the idea that unlike elastic solids, they do not support static stresses. At the
same time, it has been long known that liquids are no different to solids in supporting stresses at very high frequencies
\cite{dyre}, at times smaller than system's relaxation time. Whether the general idea of liquids being able to relax stress
can be related to slow relaxation has remained unknown.

In this paper, we propose a new framework to discuss the origin of non-exponential dynamics in a liquid. We show that, on
temperature decrease, the feed-forward interaction mechanism between local relaxation events becomes operative, resulting in
slow relaxation. This result is discussed solely on the basis of elastic response of a liquid, without invoking any
additional assumptions or postulates.

It is interesting to recall the previous discussion about the elastic aspects of liquid behaviour. Some time ago, Orowan
introduced ``condordant'' local rearrangement events \cite{orowan}. A concordant local rearrangement is accompanied by a
strain agreeing in direction with the applied external stress, and reduces the energy and local stress (see Figure 1). A
discordant rearrangement, on the other hand, increases the energy and local stress. This has led to a general result that
stress relaxation by earlier concordant events leads to the increase of stress on later relaxing regions in a system. Orowan
used this result for a heterogeneous system, in which stresses are relaxed by local rearrangements in certain regions of the
system only \cite{orowan}. Goldstein applied the same argument to a viscous liquid \cite{gold}: consider a system under
external stress which is counterbalanced by stresses supported by local regions. When a local rearrangement to a potential
minimum, biased by the external stress, occurs (a concordant event), this local region supports less stress after the event
than before; therefore, other local regions in the system should support more stress after that event than before
\cite{gold}. Noting that ``... a molecular theory of this process will be extremely complicated'', Goldstein proposed that
``the least any model of the flow process must acknowledge is that the extra stress must be supported elsewhere''
\cite{gold}. Note that this proposal implies that a supercooled liquid should be considered as an elastic medium that can
support stress relaxation and redistribution.

\begin{figure}
\begin{center}
{\scalebox{0.7}{\includegraphics{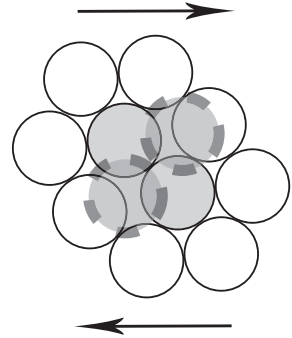}}}
\end{center}
\caption{Orowan's illustration of a concordant local rearrangement. Solid and dashed lines around the shaded atoms
correspond to initial and final positions of a rearrangement, respectively. Arrows show the direction of external stress.}
\end{figure}

We begin the discussion with introducing local relaxation events (LREs), which a system uses to relax stress locally. These
activated events have been given different names in the literature (e.g. molecular rearrangements, ``flow events'' etc
\cite{dyre}). Because each event carries a microscopic change of a macroscopic quantity (e.g. volume etc), the number of
events governs the dynamics of relaxing observables. To illustrate the argument of Orowan and Goldstein above, lets consider
the current number of LREs $n(t)$, induced in a liquid by an external perturbation. As discussed by Orowan and Goldstein,
because an external perturbation introduces bias towards concordant relaxation events, which support less stress after
relaxation, later LREs should support more stress in order to counterbalance. The increase of stress on a current relaxing
region, $\Delta p$, is therefore a monotonously increasing function of $n$.

It has been shown that activation barrier for a LRE $V$ is essentially defined by the elastic energy. In the Shoving model,
for example, $V$ is given by the work of the elastic force needed to shove aside the surrounding liquid in order for a
relaxation event to take place \cite{dyre1}. When the additional stress $\Delta p$ is loaded on a current relaxing region,
its relaxation becomes more difficult since more work is needed to shove aside the surrounding liquid (or open the cage in
Figure 1) in order to allow for relaxation event to take place. Hence the increase of $V$ is given by the increase of work
needed to overcome the additional barrier created by the elastic force due to the additional stress $\Delta p$: $\Delta
V=\int \Delta p {\rm d}q$, where the integral is taken over the reaction path of a LRE. If $q_a$ is the characteristic
activation volume \cite{dyre1}, $\Delta V=\Delta p q_a$, giving

\begin{equation}
V(n)=V_0+q_a\Delta p
\end{equation}
\noindent where $V_0$ is the initial value of the barrier.

Because $\Delta p$ is a monotonously increasing function of $n$, we find that $V(n)$ is also a monotonously increasing
function of $n$. This describes the {\it feed-forward interaction mechanism} between LREs, in that activation barriers
increase for later events.

It is important to discuss the condition under which the feed-forward interaction mechanism is operative. Let $t_{\rm s}$ be
the time needed for elastic interaction to propagate between LREs, $d$ the distance between neighbouring LREs of about 10
\AA, $c$ the speed of sound, and $\tau_0$ the oscillation period, or inverse of Debye frequency ($\tau_0=0.1$ ps). Because
$c=a/\tau_0$, where $a$ is the interatomic distance, we find that $t_{\rm s}=\tau_0 d/a$, and is temperature-independent. On
the other hand, relaxation time $\tau$, which is also the time between two consecutive relaxation events \cite{dyre},
increases on lowering the temperature. It is easy to see that at high enough temperature, when $\tau=\tau_0$, $t_{\rm
s}>\tau$ is always true because $d/a>1$. In this case, local events relax independently of each other, because at high
temperature, the time between the events is shorter than the time needed for elastic interaction to propagate between them.
Because LREs are independent, we obtain the expected high-temperature result that relaxation is exponential. On cooling the
system down, a certain temperature always gives the opposite condition: $t_{\rm s}\le \tau$. When the time between local
relaxation events exceeds the time of propagation of elastic interaction between the events, local relaxation events do not
relax independently, but ``feel'' the presence of each other.

The maximal distance between those LREs which are involved in the elastic feed-forward interaction mechanism is obtained by
setting $t_{\rm s}=\tau$:

\begin{equation}
d=a\frac{\tau}{\tau_0}
\end{equation}
\noindent where $d$ is now the range of elastic interactions in a liquid.

At high temperature, when $\tau=\tau_0$, $d$ is on the order of interatomic distances, but quickly increases on lowering the
temperature. Since $\tau=\tau_0\exp(V/kT)$, we obtain

\begin{equation}
d=a\exp\left(\frac{V}{kT}\right)
\end{equation}
\noindent where $V$ is the activation barrier. For super-Arrhneius increase of $V$, it is easy to show \cite{tau} that $d$
exceeds the system size well above $T_g$.

It follows from the above discussion that when $d$ exceeds the distance between neighbouring LRE in a liquid on the order of
10 \AA, the {\it crossover} from exponential to non-exponential relaxation takes place due to the feed-forward interaction
mechanism. The crossover temperature $T_c$ can be calculated from Eq. (3).

To calculate how $V$ depends on the current number of LREs $n$, we introduce the dynamic variable $n(t)$, the current number
of relaxing events in a sphere of radius $r=d/2$. $n(t)$ starts from zero and increases to its final value $n_{\rm r}$,
$n(t)\rightarrow n_{\rm r}$ as $t\rightarrow\infty$. Lets consider the current LRE about to relax to be in the centre of the
sphere (see Figure 2). All previous concordant LREs that are located within distance $r$ from the centre, participate in the
feed-forward interaction, increasing stress on the central region and hence the activation barrier for the central LRE. Let
$\Delta p_i(0)$ be the reduction of local stress due to a remote concordant LRE $i$. Generally, $\Delta p_i$ decays with
distance, hence we denote $\Delta p_i(r)$ as its value at distance $r$ from the centre. In what follows, we assume, for
simplicity, that $\Delta p_i(0)$ are constant, $\Delta p_i(0)=\Delta p_0$. The increase of stress on the central rearranging
region, $\Delta p$, can be calculated by integrating $\Delta p_i(r)$:

\begin{figure}
\begin{center}
{\scalebox{0.45}{\includegraphics{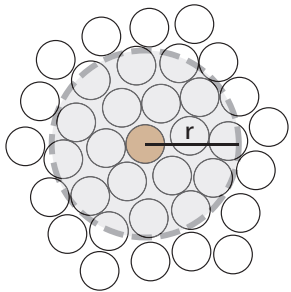}}}
\end{center}
\caption{Illustration of the feed-forward interaction mechanism between local relaxation events. This mechanism operates
within the sphere of radius $r=d/2$. Shaded and open circles represent local relaxing regions inside and outside,
respectively, of the interaction sphere.}
\end{figure}

\begin{equation}
\Delta p=\rho\int_{d_0/2}^{d/2} 4\pi r^2 \Delta p_i (r) {\rm d}r
\end{equation}

\noindent where $d_0$ is on the order of the size of a relaxing region (central region in Figure 2), $d_0\ge 3a$ (see Figure
1) and $\rho$ is the density of LREs, $\rho=6n/\pi d^3$.

It is important to note that Eq. (4) describes the redistribution of local stresses in an elastic medium. Unlike a solid, a
liquid does not support static stresses. However, it supports stresses at high frequencies in a solid-like manner
\cite{dyre}. These frequencies correspond to times smaller than relaxation time $\tau$. We introduce the distance over which
interactions in a liquid are elastic, liquid elasticity length $d_{el}$. If $c$ is the speed of sound, $d_{el}=c\tau$.
Because $c=a/\tau_0$, we find $d_{el}=a\tau/\tau_0$. As follows from Eq. (2), $d_{el}=d$. We therefore find that the
distance in which the feed-forward interaction mechanism between LREs operates is also the liquid elasticity length. Hence
we are justified to use $d/2$ as the upper limit of integration in Eq. (4).

In an elastic medium, stresses decay as $\Delta p (r)\propto 1/r^3$ \cite{elast}. Since $\Delta p(r)=\Delta p_0$ at
$r=d_0/2$, $\Delta p(r)=\Delta p_0(d_0/2r)^3$. A straightforward integration of Eq. (4), combined with Eq. (1), gives

\begin{equation}
V=V_0+V_1\frac{n}{n_r}
\end{equation}

\noindent where $V_1=\pi/2\rho_r q_a\Delta p_0 d_0^3\ln(d/d_0)$ and $\rho_r=6n_r/(\pi d^3)$ is the density of the final
number of events in the sphere.

Eq. (5) describes the feed-forward interaction mechanism in a liquid at $T<T_c$.

We are now set to derive slow relaxation. As discussed above, at high temperature above $T_c$, the feed-forward interaction
mechanism is absent, and LREs relax independently of each other. The rate of LREs, ${\rm d}n/{\rm d}t$, is proportional to
the number of unrelaxed events, $(n_{\rm r}-n)$, and the event probability, $\rho=\exp(V_0/kT)$, where $V_0$ is the
high-temperature activation barrier. Introducing $q=n/n_{\rm r}$, we write:

\begin{equation}
\frac{{\rm d}q}{{\rm d}t}=\exp\left(-\frac{V_0}{kT}\right)(1-q)
\end{equation}

\noindent where $t$ is dimensionless time $t\rightarrow t/\tau_0$. The solution of Eq. (6) is the expected high-temperature
exponential relaxation, $q\equiv n/n_{\rm r}=1-\exp(-{t/\tau})$, where $\tau=\tau_0\exp(V_0/kT)$.

As temperature is lowered below $T_c$, the feed-forward interaction mechanism between consecutive LREs becomes operative as
discussed above. Because this mechanism is described by Eq. (5), the event probability $\rho$ becomes dependent on $n$ (or
$q$), and the rate equation is (compare with Eq. (6):

\begin{equation}
\frac{{\rm d}q}{{\rm d}t}=\exp\left(-\frac{V_{0}}{kT}\right)(1-q) \exp(-\alpha q)
\end{equation}
\noindent where $\alpha=V_1/kT$.

Unlike Eq. (6), Eq. (7) describes slow relaxation due to the factor $\exp(-\alpha q)$. Non-exponential relaxation becomes
exponential at high temperature when $\alpha$ is small. Since SER is one of the simplest functions that describes slow
relaxation, we attempt to fit the solution of Eq. (7) to $q=1-\exp{(-(t/\tau)^\beta})$. Because the degree of
non-exponentiality of the solution of Eq. (7) is controlled by $\alpha$ only, we set $\exp(-V_0/kT)=1$. This affects $\tau$
but not $\beta$. In this form, Eq. (7) is a one-parameter description of non-exponential relaxation. We solve Eq. (7)
numerically for different values of $\alpha$, and fit the solution to SER using the least-squares method. A reasonably good
quality of the fits is illustrated in Figure 3 over about 5 decades of reduced time. Note that smaller temperature (larger
$\alpha$) gives smaller $\beta$ (see the legend in Figure 3), consistent with experimental results \cite{ngai,phillips}.

\begin{figure}
\begin{center}
{\scalebox{0.8}{\includegraphics{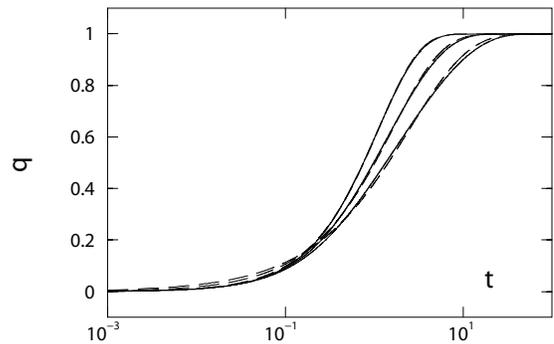}}}
\end{center}
\caption{Solid lines are solutions of Eq. (7), dashed lines are fits to SER. From left to right, $\alpha$=0.3, 1.2 and 2.2.
SER parameters ($\beta$, $\tau$) are (0.94, 1.12), (0.78, 1.62) and (0.66, 2.51) in the same order.}
\end{figure}

It is interesting to note that on lowering the temperature, we expect to find the crossover of $\beta$ to the lower slope.
Using Eq. (3), we find $V_1\propto \ln(d/d_0)=(\ln(a/d_0)+V/kT)$. This relation remains true as long as, on lowering the
temperature, $d<L$, where $L$ is the system size. For propylene carbonate, for example, a simple calculation shows that
$d=L$ takes place at temperature above $T_g$ \cite{tau}. When $d\ge L$, the upper limit of integral in Eq. (4) is $L$, and
$V_1\propto \ln(L/d_0)$, and is temperature-independent. Hence at $d=L$, $V_1$ shows a kink and starts to saturate to a
constant value. Because $\beta$ decreases with $\alpha=V_1/kT$ (see Eq. (7)), we find that $d=L$ should mark the crossover
of $\beta$ to the lower slope. This is indeed what is seen in several glass-forming systems at low temperature \cite{casa}.

We also note that the proposed picture of slow relaxation gives rise to the property that has been discussed quite intensely
recently, namely that relaxation in supercooled liquids and glasses is dynamically heterogeneous \cite{hetero}. Dynamic
heterogeneity naturally arises in our picture, since activation barriers increase for later LREs that are located in
different regions of the system. This gives different waiting times in different regions of the system and dynamic
heterogeneity.

In this paper, our main intention has been to identify and discuss the origin of slow relaxation in a liquid. We proposed
that slow relaxation can be understood solely on the basis of liquid elasticity, without invoking any additional assumptions
or postulates. We have shown that on lowering the temperature, the feed-forward interaction mechanism between LREs becomes
operative and sets slow relaxation. We have offered a formalism to describe this mechanism. The mathematical treatment can
be developed further in future work. Here we note that it is an encouraging sign that we can discuss slow relaxation in a
liquid using only concepts from elasticity, one of the most robust areas of physics.

Compared with other models of slow relaxation, our picture is probably the closest in its spirit to the Coupling model (CM)
of the glass transition \cite{ngai}. The main assumption of the CM model is the postulate that non-exponentiality is the
result of some sort of cooperativity of relaxation that kicks in on lowering the temperature. Here, we discussed that the
origin of cooperativity is the feed-forward interaction mechanism between LREs. Absent at high temperature, it becomes
operative on lowering the temperature, setting the cooperative character of relaxation and slow dynamics.

We have recently shown that the feed-forward interaction mechanism, which gives non-exponential relaxation, is also
responsible for the Vogel-Fulcher-Tamman (VFT) law for relaxation time. Here, super-Arrhenius behaviour is associated with
increase, on lowering the temperature, of the range of the feed-forward interaction \cite{tau}. Joining this result with the
present discussion, we can conclude that the two most important open questions in the area of glass transition, the origin
of SER and the VFT law \cite{dyre}, can be understood on the basis of elasticity of a liquid.

I am grateful to V. V. Brazhkin, J. C. Dyre and M. Goldstein for discussions, and to EPSRC for support.

\end{document}